\documentstyle[11pt,moriond,epsfig]{article}

\def\ETAL{{\em et al.}}

\def\be{\begin{equation}}
\def\ee{\end{equation}}
\def\bea{\begin{eqnarray}}
\def\eea{\end{eqnarray}}

\begin{document}
\vspace*{4cm}
\title{Charm Production at NuTeV}

\author{ 
 T.~Adams$^{4,*}$, {A.~Alton$^{4}$}, {S.~Avvakumov$^{7}$},
 L.~de~Barbaro$^{5}$, P.~de~Barbaro$^{7}$, R.~H.~Bernstein$^{3}$,
 A.~Bodek$^{7}$, T.~Bolton$^{4}$, J.~Brau$^{6}$, D.~Buchholz$^{5}$,
 H.~Budd$^{7}$, L.~Bugel$^{3}$, J.~Conrad$^{2}$, R.~B.~Drucker$^{6}$,
 R.~Frey$^{6}$, {J.~Formaggio$^{2}$},{J.~Goldman$^{4}$}, 
 {M.~Goncharov$^{4}$},
 D.~A.~Harris$^{7}$, R.~A.~Johnson$^{1}$, S.~Koutsoliotas$^{2}$,
 {J.~H.~Kim$^{2}$}, M.~J.~Lamm$^{3}$, W.~Marsh$^{3}$, 
 {D.~Mason$^{6}$}, {C.~McNulty$^{2}$}, K.~S.~McFarland$^{7}$, 
 D.~Naples$^{4}$, P.~Nienaber$^{3}$, {A.~Romosan$^{2}$}, W.~K.~Sakumoto$^{7}$,
 H.~Schellman$^{5}$, M.~H.~Shaevitz$^{2}$, P.~Spentzouris$^{2}$ , 
 E.~G.~Stern$^{2}$, {B.~Tamminga$^{2}$}, {M.~Vakili$^{1}$},
 {A.~Vaitaitis$^{2}$}, {V.~Wu$^{1}$}, {U.~K.~Yang$^{7}$}, J.~Yu$^{3}$ and 
 {G.~P.~Zeller$^{5}$}}

\address{
 $^{*}$Presented by T. Adams \\
 $^{1}$University of Cincinnati, Cincinnati, OH, USA \\            
 $^{2}$Columbia University, New York, NY, USA \\                   
 $^{3}$Fermi National Accelerator Laboratory, Batavia, IL, USA \\  
 $^{4}$Kansas State University, Manhattan, KS, USA \\              
 $^{5}$Northwestern University, Evanston, IL, USA \\               
 $^{6}$University of Oregon, Eugene, OR, USA \\                    
 $^{7}$University of Rochester, Rochester, NY, USA                 
}


\maketitle\abstracts{
Neutrino deep-inelastic scattering provides a means to study both the
strange and charm content of the nucleon.  The NuTeV experiment 
(Fermilab E-815) takes full advantage of separated neutrino and 
anti-neutrino beams to probe the nucleon.  
The strange sea is studied
with charged-current charm production resulting in an opposite-signed
two muon final state.  The charm content of the nucleon is probed via
neutral-current charm production creating an event with a single
wrong-signed muon.  Preliminary results are presented for both
analyses.}

\begin{figure}
 \hspace{1cm} \psfig{figure=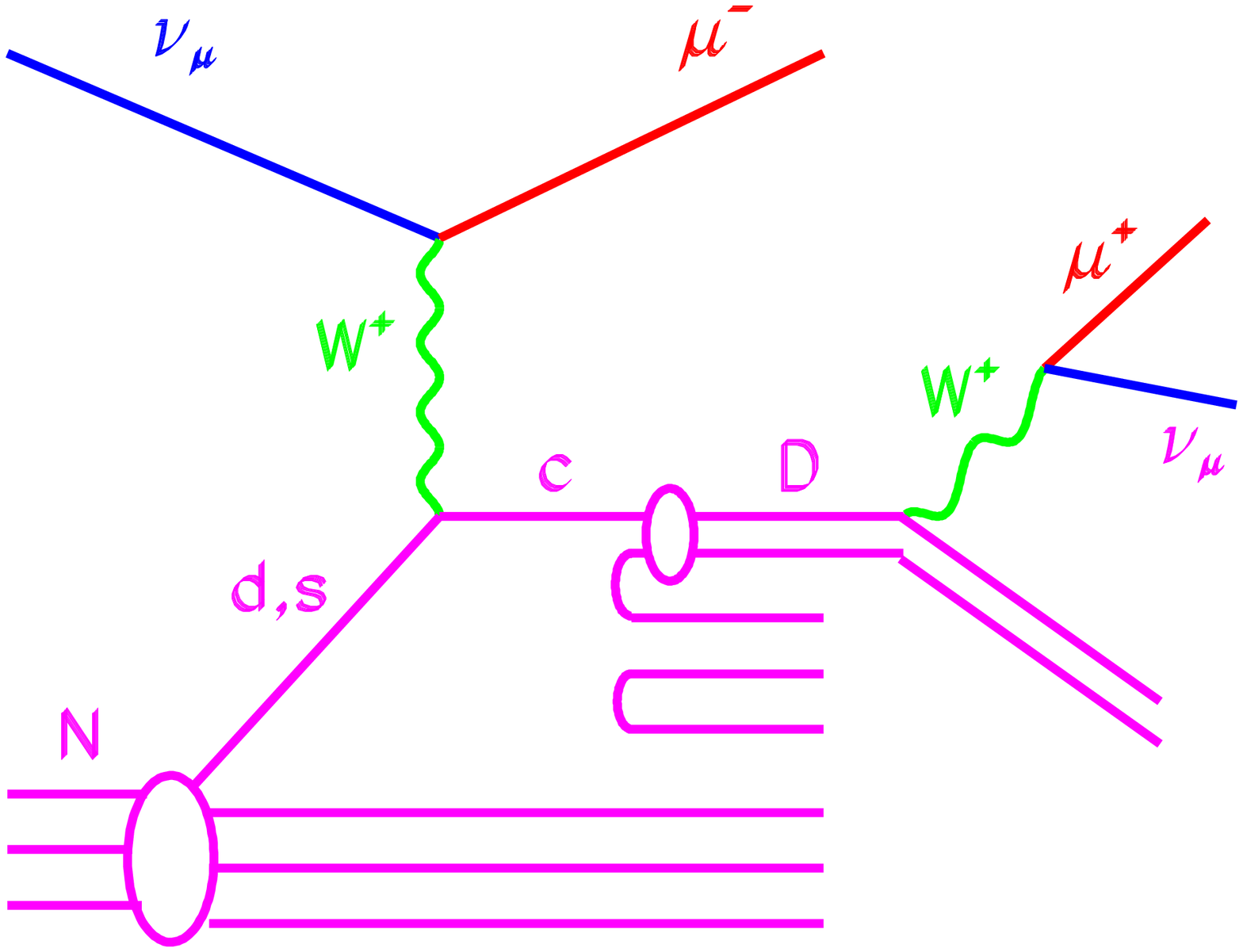,width=6cm} \hfill
 \psfig{figure=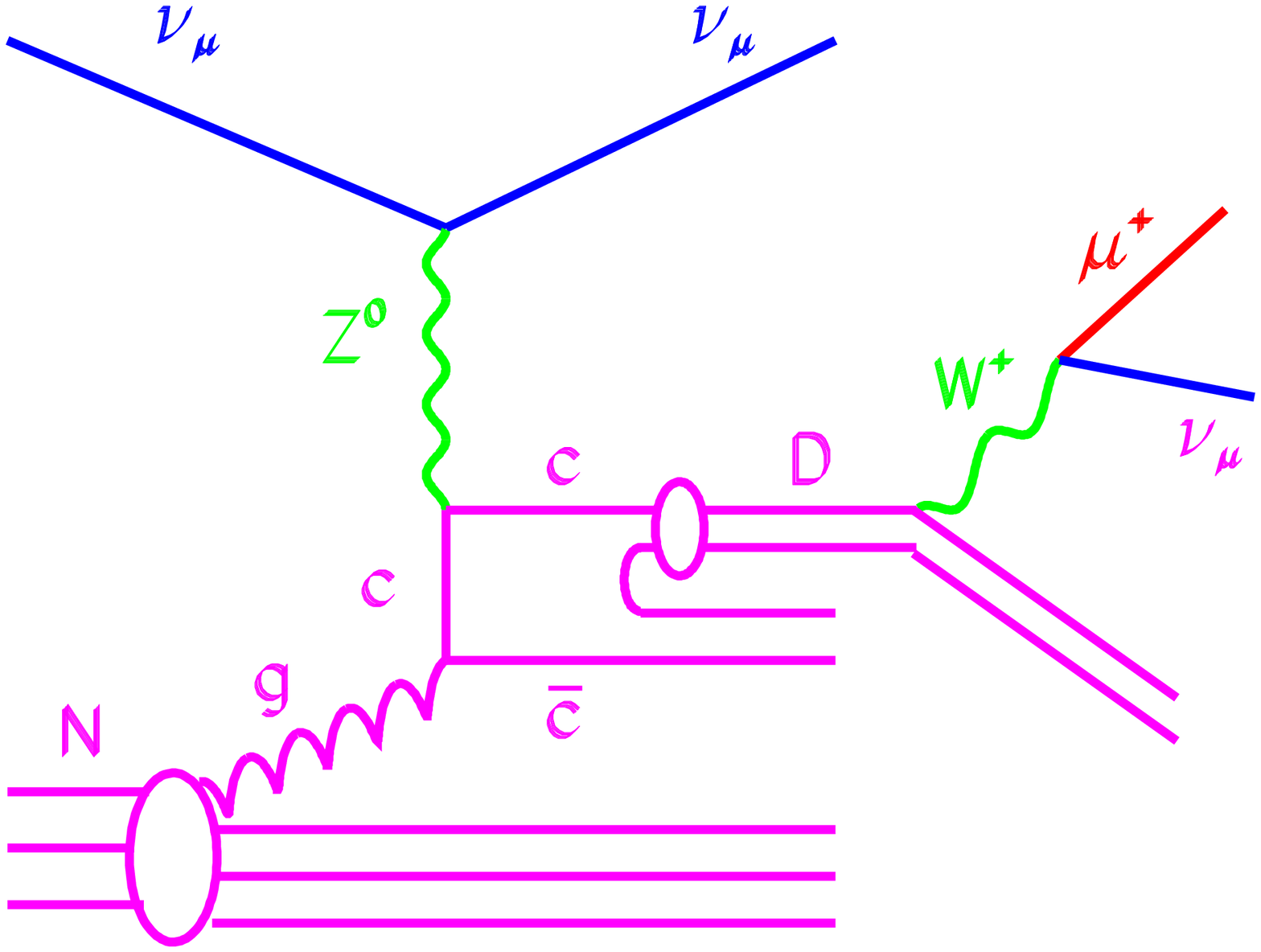,width=6cm} \hspace{1cm}

 \hspace*{3cm} (a) \hfill (b) \hspace{4cm}
 \caption{Feynman diagrams for (a) charged-current and (b) neutral-current charm
production via neutrino DIS.
 \label{fig:tadams:feyn_diagrams}}
\end{figure}

Charm production is a significant fraction of the total neutrino
deep-inelastic scattering (DIS) cross-section.  The semi-muonic
decay of charm mesons creates unique final states to study exclusive
charm production.  For charged-current reactions, an opposite-signed
two muon (dimuon) final state is available.  Neutral-current production
can create an event with a single muon with charge opposite that
expected from a charged-current interaction (wrong-signed
muon).  Feynman diagrams for both reactions are shown in 
Fig.~\ref{fig:tadams:feyn_diagrams}.  

The NuTeV experiment studies $\nu$ and $\bar{\nu}$ DIS
using the Sign-Selected Quadrupole Train (SSQT) beam and the
Fermilab Lab E detector.\cite{bib:tadams:labe}  The SSQT allows 
separate running
of neutrino and anti-neutrino beams.  The (anti-)neutrino beam is incident
upon the Lab E target/calorimeter which has 42 segments each consisting
of two liquid-scintillator counters and a drift chamber interspersed
with 20 cm of iron.  The calorimeter provides energy and position measurement 
for hadronic/electromagnetic showers and deeply-penetrating muon tracks.
Immediately downstream is a toroid spectrometer which provides 
momentum measurement for muons.  Analyses presented here
use the full NuTeV data sample from the Fermilab 1996-97 fixed 
target run.

\section{Charged-Current Charm Production \label{sec:tadams:cccharm}}

Neutrino charged-current DIS charm production results from scattering off
$d$ or $s$ quarks in the nucleon.  The Cabibbo suppression of 
scattering off $d$ quarks greatly enhances the contribution of
the $s$ quarks.
This allows the probing of the strange content of the nucleon.


The dimuon data set consists of events passing fiducial and kinematic
selections and containing a hadronic shower ($>$ 10 GeV) with at least 
two muons.  One muon must be toroid analyzed with more than 9 GeV in
energy while the other muon must have more than 5 GeV in energy.
The two sources of such events are charged-current charm production and a
charged-current event with a $\pi/K$ decaying within the
shower.  


Data and Monte Carlo are binned in three variables: $x_{vis} = 
\frac{E_{vis} E_{\mu 1} \theta^2_{\mu 1}}{2 m_p (E_{\mu 2} + E_{had})}$, 
$E_{vis} = E_{\mu 1} + E_{\mu 2} + E_{had}$, and 
$z_{vis} = \frac{E_{\mu 2}}{E_{\mu 2} + E_{had}}$ where the
subscript $vis$ ($visible$) refers to measured variables, $\mu 1$ is
the primary muon and $\mu 2$ is the secondary muon.  The 
Monte Carlo is weighted by the leading-order (LO) cross-section and
normalized to agree with the two-muon data.

Four parameters are used to describe charged-current charm
production.
$\kappa$ determines the size of the strange sea relative to the
non-strange sea ($\sim \frac{2\bar{s}}{\bar{u}+\bar{d}}$).   The shape
of the strange sea is described by $(1-x)^\beta$.  $m_c$ is the
mass of the charm quark and $\epsilon$ determines the 
Collins-Spiller fragmentation.~\cite{bib:tadams:collins}
These parameters are varied to find the best agreement between data
and Monte Carlo.
The preliminary results of the fit are:
\bea
 \kappa   & = & 0.42 \pm 0.07 \pm 0.06 \nonumber \\
 \beta    & = & 8.5 \pm 0.56 \pm 0.39 \hspace{0.5cm} (Q^2=16~GeV^2) \nonumber \\
 m_c      & = & 1.24 \pm 0.25 \pm 0.46~~GeV \nonumber \\
 \epsilon & = & 0.93 \pm 0.11 \pm 0.15 \nonumber
\eea
where the first error is statistical and the second error is 
systematic.  The largest systematic error is from the flux which is
expected to be reduced by further work.  However, the errors for
$\kappa$ and $\alpha$ are already statistics limited so the improvement
will be primarily for $m_c$ and $\epsilon$.  

\begin{figure}
 \psfig{figure=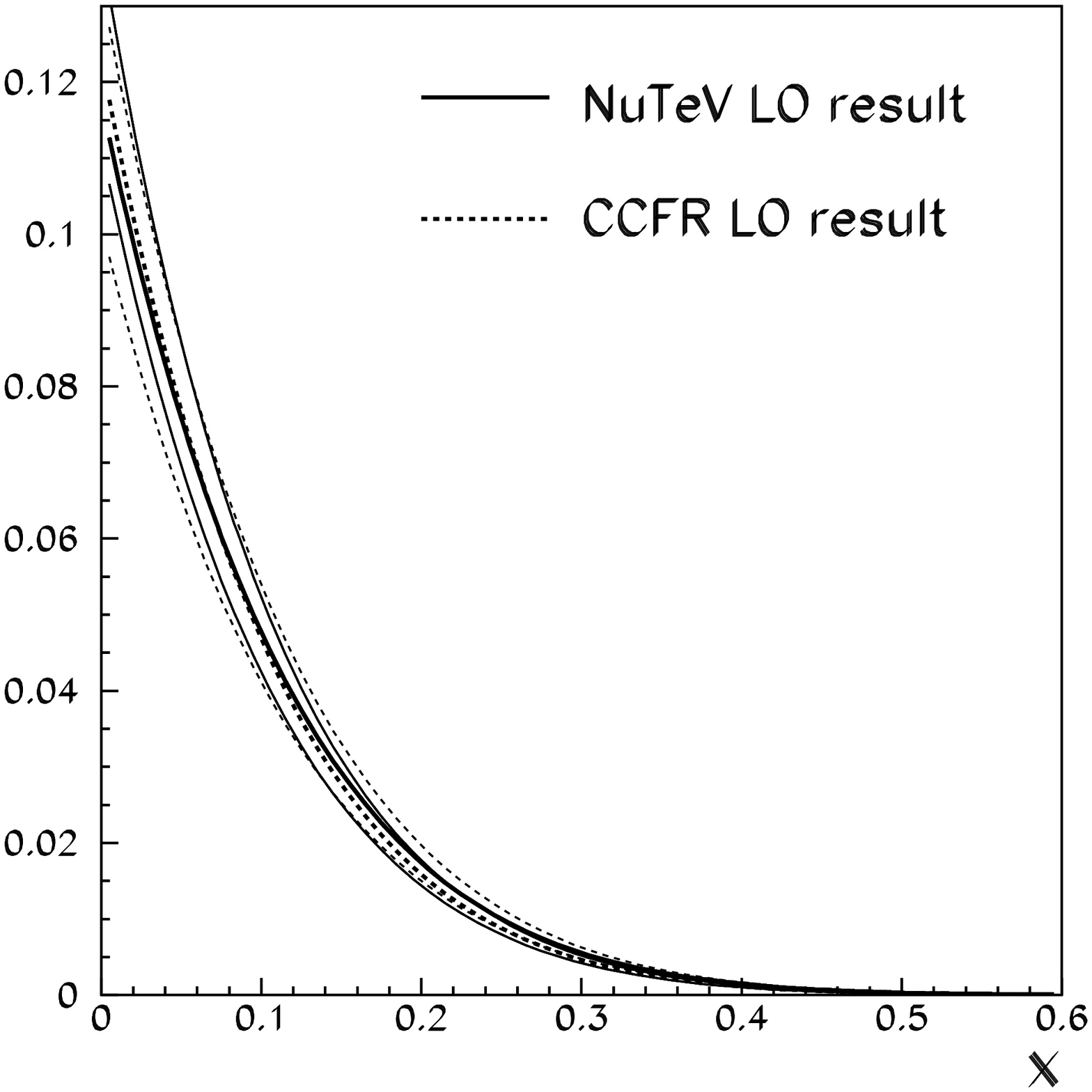,width=7.5cm} \hfill
 \psfig{figure=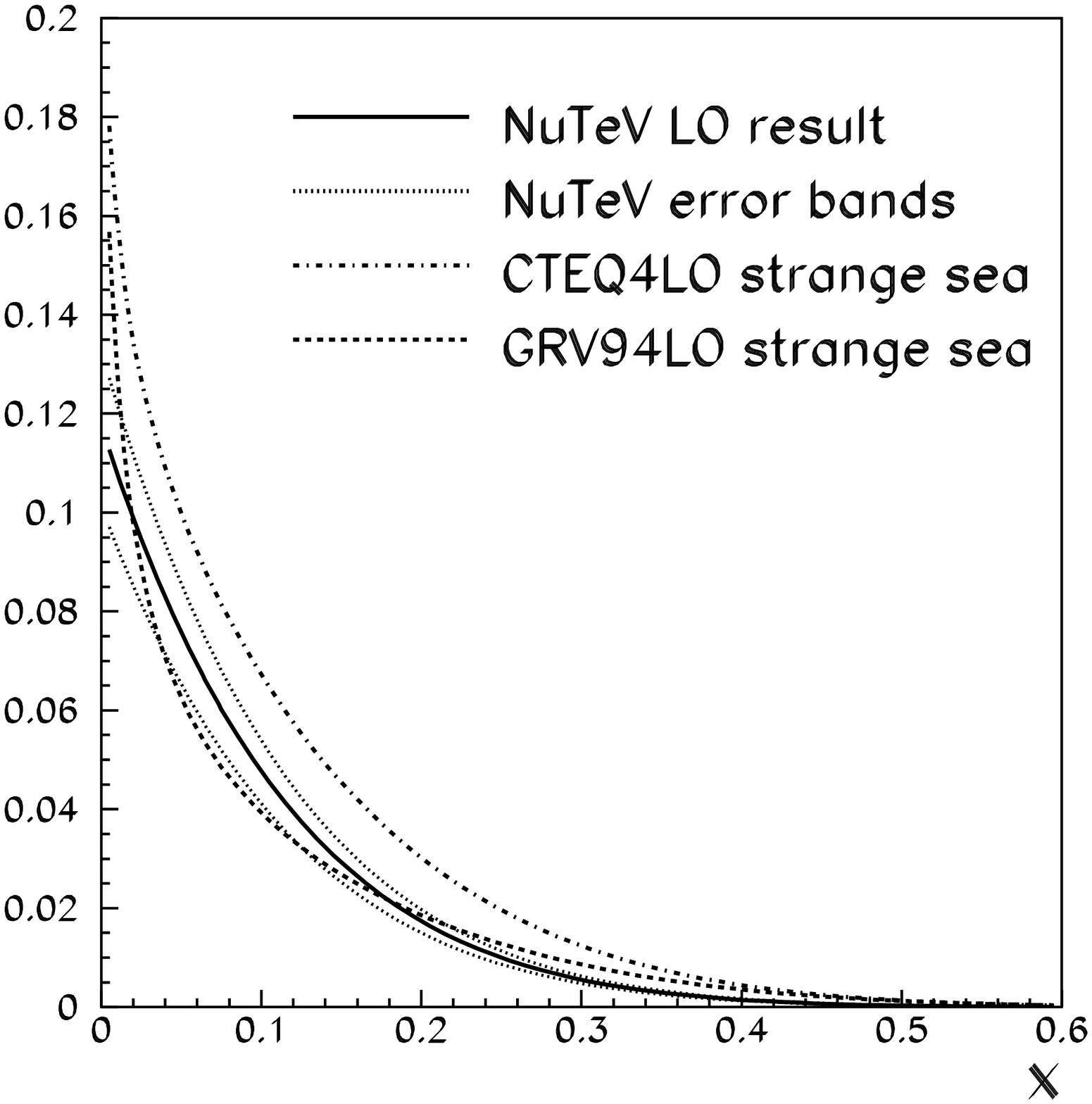,width=7.5cm}

 \hspace*{4cm} (a) \hfill (b) \hspace{3cm}
 \caption{Comparison of the NuTeV (with error bands) result to: 
CCFR result (a); theoretical predictions from CTEQ and GRV.
 \label{fig:tadams:ssea}}
\end{figure}

The results of this analysis for the strange sea compare favorably with 
previous measurements.  CHARM II,~\cite{bib:tadams:charmii}
CCFR~\cite{bib:tadams:rabin} and CDHS~\cite{bib:tadams:cdhs} have all
measured the size of the strange sea to be $\sim40\%$ of the non-strange
sea.  
Figure~\ref{fig:tadams:ssea}(a) shows
a comparison of the results of the CCFR and NuTeV strange sea measurements.
There is excellent agreement between the two measurements.
Figure~\ref{fig:tadams:ssea}(b) compares the NuTeV result to theoretical
predictions from CTEQ and GRV.  The NuTeV result is lower than the
CTEQ prediction while in better agreement with GRV.

\section{Neutral-Current Charm Production \label{sec:tadams:nccharm}}

Neutral-current (NC) charm production occurs via scattering off charm
quarks in the nucleon (Fig.~\ref{fig:tadams:feyn_diagrams}(b)).  Analysis
of this reaction probes the charm content of the nucleon.  While some
models allow for an intrinsic charm content of the nucleon,
this analysis only considers gluon splitting as its source for charm.

This analysis considers events which have a single muon with
the opposite charge from charged-current interactions of the selected
beam type (wrong-sign events).  This is possible because of NuTeV's 
ability to select
between neutrino and anti-neutrino beams and the low contamination
of wrong type neutrinos.

There are three sources of wrong-sign events other than NC charm
production.  The largest is charged-current impurities including
events from anti-neutrinos in neutrino mode.
This contamination is less than $2\times10^{-3}$ of the
beam intensity.  Also included in this category are 
charged-current events where the sign of the muon is mis-identified
(generally due to a large scatter in the toroid spectrometer).  The
second largest source comes from two-muon events (see 
Section~\ref{sec:tadams:cccharm}) where the primary muon is not
measured because it is of low energy or exits the detector.  The
third source comes from NC events where a $\pi/K$ in the shower decays prior
to interacting.

\begin{figure}
 \psfig{figure=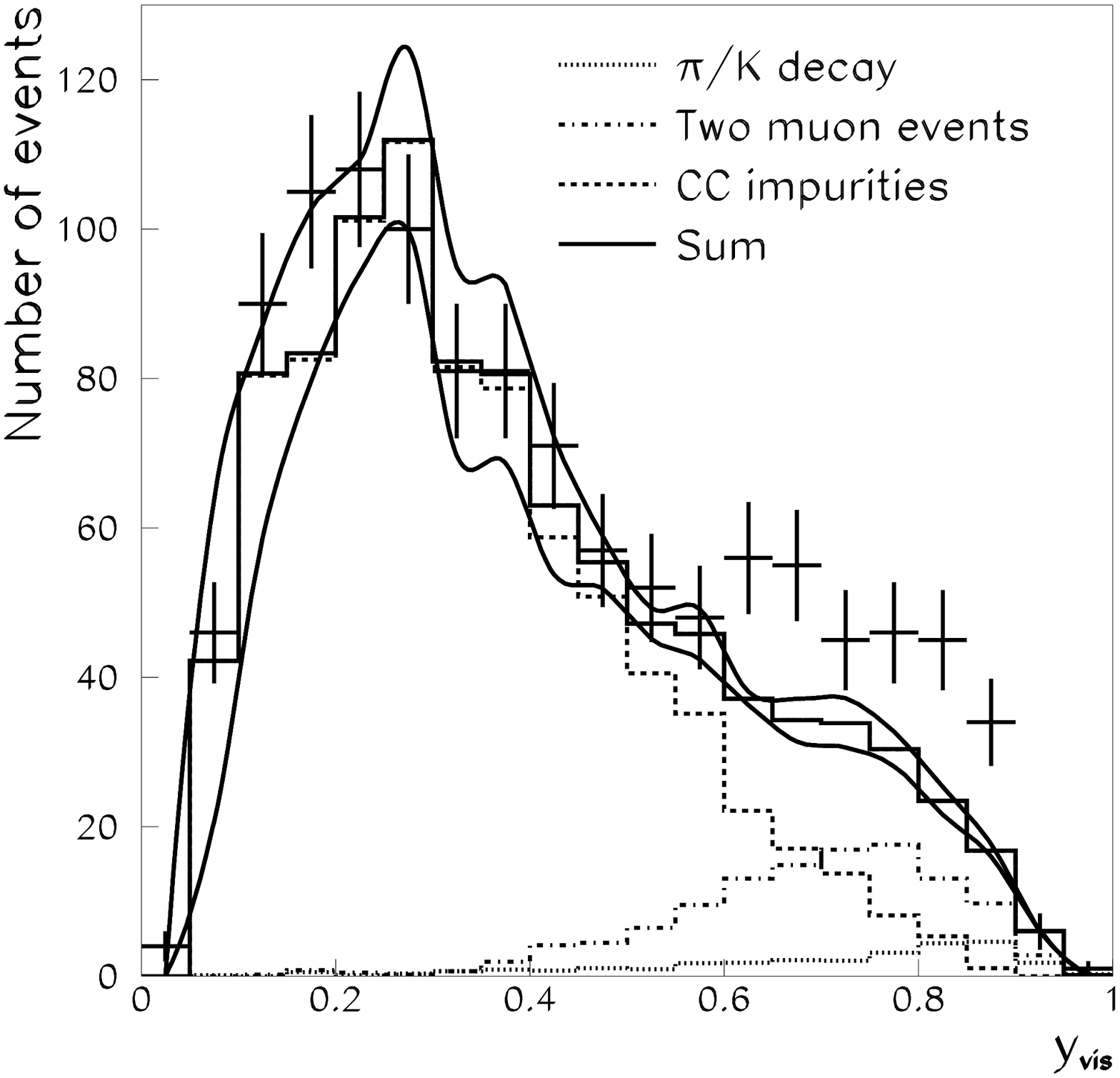,width=7.5cm} \hfill
 \psfig{figure=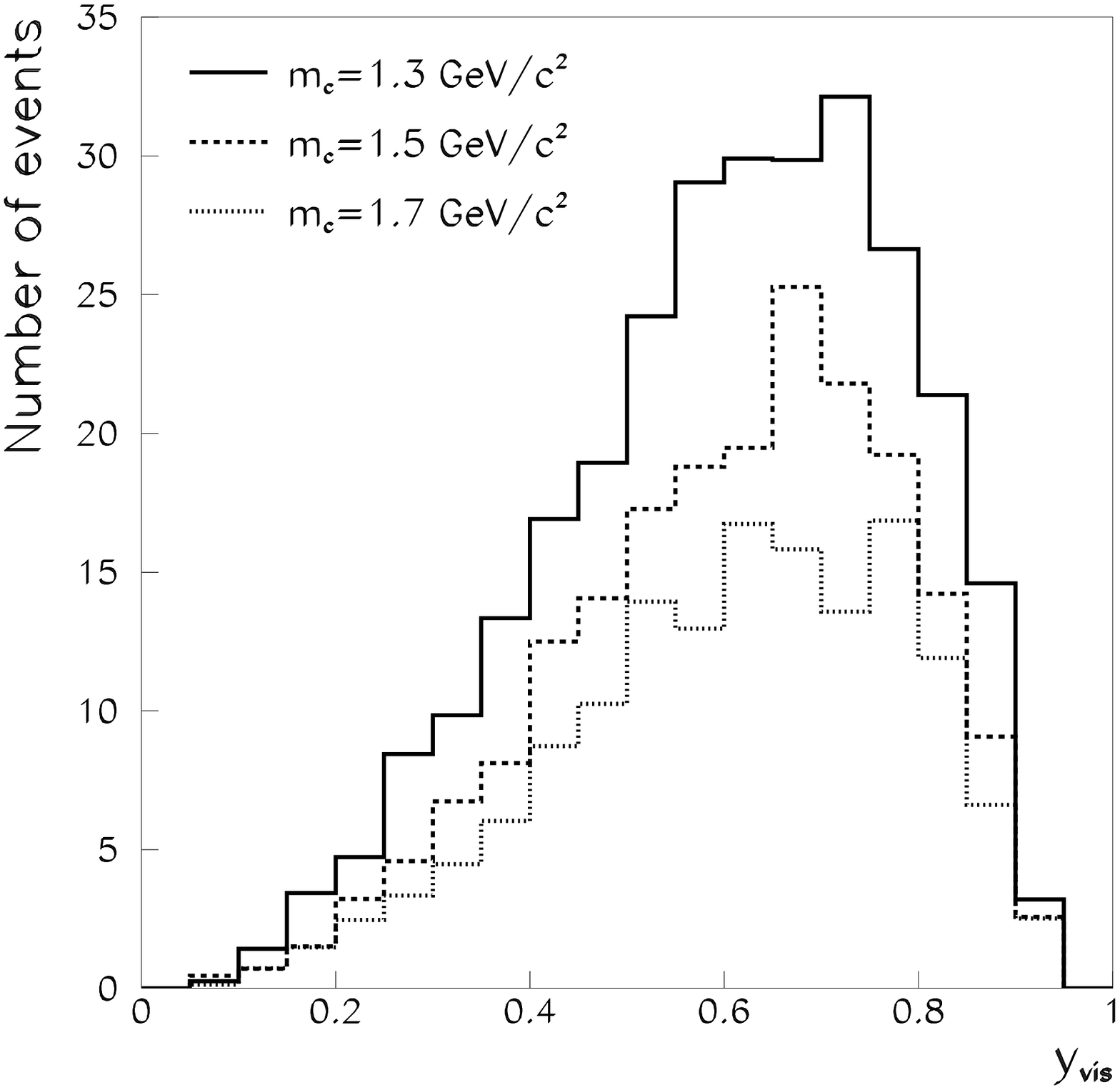,width=7.5cm} 

 \hspace*{4cm} (a) \hfill (b) \hspace{3cm}
 \caption{Neutral-current charm production ($\nu$ mode only).  
(a) NuTeV wrong-sign muon
data (points) compared to the sum of the background sources (solid histogram)
and systematic errors (curve).  The individual background components are
also shown as histograms.; (b)  Predictions of neutral-current charm
production for various values of $m_c$ = 1.3,1.5,1.7.
 \label{fig:tadams:wsm}}
\end{figure}

Figure~\ref{fig:tadams:wsm}(a) shows the wrong-signed muon data from
NuTeV (points) for the variable $y_{vis} = \frac{E_{HAD}}{E_{HAD} + E_\mu}$.  
The predictions for the individual background sources and
their sum are also shown.  A clear excess of events is seen at 
high $y_{vis}$.
Figure~\ref{fig:tadams:wsm}(b) shows predictions for neutral-current
charm production which peaks in the same region as the 
excess.~\cite{bib:tadams:nccharm}

\section{Summary}

The NuTeV sign-selected neutrino beams allows for improved studies of
the strange and charm contributions to the nucleon sea.  The strange sea
is observed to be in agreement with previous measurements with a size
which is $0.42 \pm 0.07 \pm 0.06$ the size of the average non-strange sea and
a shape $(1-x)^{8.5 \pm 0.56 \pm 0.39}$ (at $Q^2=16$ GeV$^2$).  The 
wrong-signed muon data shows a clear excess of events which is
consistent with neutral-current charm production.  Additional details
of both analyses are available.~\cite{bib:tadams:epic99}

\section*{Acknowledgments}

We gratefully acknowledge the staff of Fermilab 
for their significant contributions to this work.

\end{document}